# A rheological state diagram for rough colloids in shear flow

Lilian C. Hsiao[1,*], Safa Jamali[2], Daniel J. Beltran-Villegas[3], Emmanouil Glynos[4,§],

Peter F. Green[4], Ronald G. Larson[3] & Michael J. Solomon[3]

[1] Department of Chemical and Biomolecular Engineering, North Carolina State University, Raleigh, NC 27695, USA.

[2] Department of Chemical Engineering, Massachusetts Institute of Technology, Cambridge, MA 02139, USA.

[3] Department of Chemical Engineering, University of Michigan, Ann Arbor, MI 48109, USA.

[4] Department of Material Science and Engineering, University of Michigan, Ann Arbor, MI 48109, USA.

[§] Present address: Institute of Electronic Structure and Laser, Foundation for Research and Technology-Hellas, Crete, Greece.

[*] Corresponding author. e-mail: lilian_hsiao@ncsu.edu

**The flow of dense suspensions, glasses, and granular materials is heavily influenced by frictional interactions between constituent particles. Although this friction is related to particle roughness and geometry, their connection is not straightforward. To date, neither hydrodynamics nor friction has successfully explained the full range of flow phenomena in concentrated suspensions. Particles with asperities represent a case in point. When particulate materials flow, hydrodynamic and contact stresses give rise to continuous and discontinuous shear thickening,[1-3] dilatancy,[4] and shear jamming.[5-7] Lubrication hydrodynamics predict that particle roughness induces fore-aft asymmetry in the flow field,[8,9] but fail to completely capture two key rheological properties – namely, that the viscosity increases drastically and the first normal stress difference can switch signs as volume fraction increases. Presumably, this discrepancy is because solid-solid contacts become prevalent. Yet, simulations that account for interparticle friction are also unable to fully predict these properties. Furthermore, experiments show that rheological behavior can vary depending on particle roughness and deformability.[10] Thus, there is a need to develop a general framework that connects the effects of particle geometry to flow behavior, given the recent interest in materials that can turn solid-like on demand[11-13] and the ubiquitous processing of particulate suspensions.[3]**

The physicist and engineer Osborne Reynolds modeled frictionless balls using a bag of round shotgun projectiles, and described dilatancy as the expansion of volume required for rigid particles in close contact to rearrange during flow.[14] This expansion occurs because particles cannot easily find direct flow paths within the confined environment. Indeed, many types of modern suspensions undergo both dilatancy and shear thickening at high volume fraction ($\phi$), and friction is known to affect both of these flow phenomena. While dilatancy is characterized by positive first normal stress differences ($N_1 > 0$) due to volume expansion against boundaries, shear thickening is a related but independent phenomenon defined by an increase in the suspension viscosity $\eta$ at large shear stresses ($\sigma$) or shear rates ($\dot{\gamma}$). The dilatant and shear thickening nature of concentrated suspensions is famously exemplified by the ability of cornstarch pools to support the weight of people running across them, and is typically avoided in material processing due to issues with clogging or catastrophic failure.[3] When suspensions shear thicken at high $\phi$, it is frequently accompanied by complex flow behavior that includes shear banding,[15, 16] slip,[17] and slow stress decays.[18] The degree of shear thickening can gradually range from a few-fold increase to orders of magnitude increase in $\eta$ as a function of $\sigma$. These distinctions are commonly used as working definitions for continuous shear thickening (CST) and discontinuous shear thickening (DST) in the experimental literature.[19] We maintain the same terminology for consistency, and demarcate CST and DST using the shear thickening power $\beta$ as the slope of log($\eta$) plotted against log($\sigma$), where CST occurs at $0.1 \leq \beta \leq 0.65$ and DST occurs at $0.65 < \beta \leq 1.0$. These categories are a convenient classification of the magnitude of the rheological response rather than the identification of a fundamental physical transition.

Based on the above observations, particle simulations incorporating both lubrication forces and frictional contacts have been recently developed.[20, 21] Recent simulations show that the tangential nature of friction is necessary to produce the large jumps in $\eta$ and positive $N_1$ values sometimes seen in DST,[21] with some suggesting that frictional stresses could dominate even in the regime of CST.[22] This is in contrast to an extensive body of work that connects CST to the formation of hydroclusters lubricated by thin fluid layers.[3, 23] Such hydroclusters result in negative $N_1$ values at large $\sigma$ due to anisotropies in the shear-induced microstructure.[24, 25] Experiments performed with suspensions of smooth particles show that $N_1$ is almost always negative in both CST and DST regimes.[26] Dilatant flows have been reported in particles with surface asperities[27] or at very high $\phi$.[28]

The reasons for the discrepancies between theory, simulations, and even amongst experiments are unclear, particularly when particles that are neither spherical nor smooth are encountered. Few studies quantitatively relate particle surface roughness to frictional effects in the flow of their suspensions, resulting in a lack of consensus amongst different researchers. On one hand, some find that as $\phi$ and $\sigma$ increase, large frictional stresses give rise to shear thickening at high shear rates.[5, 15, 20, 22, 28-30] Others find that lubrication combined with particle deformation can give rise to DST with negative $N_1$ values, and that contact forces may be unimportant in most cases.[10, 23, 25] We seek to resolve these apparent contradictions by systematically tuning the roughness of model colloids, investigating $\eta$ and $N_1$ under steady shear, and finally demonstrating how surface roughness influences the transition between shear thickening and dilatancy.

Our system uses density and refractive index-matched colloidal dispersions that contain particles in the size range of $2a \sim 1.9$-$2.5$ μm, for which shear thickening occurs without inertial contributions. We find that these dense suspensions of rough colloids shear thicken and dilate at unusually low volume fractions and shear rates when the root-mean-squared (RMS) roughness is larger than the lubrication length scale. Our experiments and simulations show that lubrication is indeed dominant in moderately concentrated suspensions. Frictional interactions are likely to become important, however, when particles are pushed into contact at high shear rates, volume fractions, or surface roughness. A surprising finding is that the onset stress for shear thickening is independent of $\phi$ only in the case of smooth particles, and that roughness decreases this onset stress by reducing the force required to push colloids into contact by orders of magnitude. The results are presented as a rheological state diagram that provides insight into the transition from shear thickening to dilatant regimes for colloidal suspensions.

Experiments are performed using poly(methyl methacrylate) (PMMA) colloids sterically stabilized by a 10-nm layer of poly(12-hydroxystearic acid) (PHSA)[31] (**Fig. 1a**) dispersed in an organic solvent designed to minimize sedimentation and van der Waals forces. The RMS roughness of the colloids is tuned by varying the concentration of a crosslinker molecule with respect to the monomer. The crosslinker is thought to induce heterogeneity during oligomer precipitation, resulting in size-monodisperse rough particles when the reaction is complete (**Fig. S1**, **Table S1**).[32] Using this method, we synthesize PMMA colloids with four classes of surface asperities: smooth (SM), slightly rough (SL), medium rough (MR), and very rough (VR), and quantify their topographic profiles using an atomic force microscope (AFM) in tapping mode (**Fig. 1b**). The 3D topography data is fitted to a sphere with an effective radius, $a_{\text{eff}}$, where the deviation of surface profiles from $a_{\text{eff}}$ is minimized by a least squares fitting. Volume fractions are computed using $a_{\text{eff}}$ values from AFM imaging and from 3D image volumes of particle suspensions captured using confocal laser scanning microscopy (CLSM). We then apply the relation $\phi = \frac{4}{3}\pi a_{\text{eff}}^3 / V_{box}$, where $V_{box}$ is the total volume of the CLSM image box that is analyzed. We find that our experimental method of weighing particles and solvent yields the correct value of ϕ to 2% (**Fig. S2**). An additional uncertainty in ϕ of 2% is present due to particles swelling over time in the solvent (**Fig. S3**). We characterize roughness using the autocovariance of the topographic profile[33] in spherical coordinates,

$B(\psi) = \frac{1}{N} \sum_{\substack{i,j=1 \\ i \neq j}}^{\psi} (|\mathbf{r}_i| - a_{eff})(|\mathbf{r}_j| - a_{eff})$, where $N$ is the total number of data points analyzed

and $\psi = \cos^{-1}\left(\frac{\mathbf{r}_i \cdot \mathbf{r}_j}{|\mathbf{r}_i||\mathbf{r}_j|}\right)$ is the angle between two vectors $\mathbf{r}_i$ and $\mathbf{r}_j$ as defined in **Fig. 1c**.

The relative RMS roughness is given by $(B(\psi = 0)/a_{eff}^2)^{1/2}$ to account for different particle sizes. The full range of our experimental conditions is provided in **Table S2**.

Colloids with large surface roughness shear thicken more readily. When vials containing suspensions at ϕ = 0.52 are inverted, smooth colloids flow like a viscous fluid whereas rough colloids shear thicken and form finger-like structures under gravitational stress (**Fig. 1d**, **Movie S1**). To fully quantify the shear thickening as a function of roughness and ϕ, we measure $\eta$ and $N_1$ as a function of $\sigma$ in a stress-controlled rheometer. Performing stress sweeps up and down shows that the flow is completely reversible for SM colloids at ϕ = 0.55, whereas MR colloids show hysteresis in $\eta(\sigma)$ and $N_1(\sigma)$ at ϕ = 0.535 even when steady state flow conditions are imposed (**Fig. S4**). We also verify the absence of global slip by comparing the flow curve of MR colloids with that collected with a different cone-and-plate geometry (**Fig. S5**).

Flow curves of the shear thickening colloidal suspensions are presented in **Fig. 2**. At low $\sigma$, the suspensions flow with a nearly constant relative shear viscosity, $\eta_{r,N}$. As the stress is further increased, $\eta$ begins to increase significantly at the point of shear thickening. We define the onset stress, $\sigma_c$, by the intersection of power laws fitted to the Newtonian and shear thickening regimes of the flow curves (**Fig. 2a**). SM colloids undergo CST in the range of $\phi$ that are tested, with a gradual progression towards DST in all other types of particles with surface roughness (SL colloids at $\phi = 0.55$, MR colloids at $\phi \geq 0.50$, and VR colloids at $\phi \geq 0.45$). Although $\sigma_c$ is independent of $\phi$ for SM colloids, we find that $\sigma_c$ decreases with increasing $\phi$ for SL, MR, and VR colloids. These observations for rough particles are markedly different from other experiments of smooth particles: multiple studies show that $\sigma_c$ is typically constant in the regimes of CST and DST.[17, 28, 30] We address the rationale for the difference between smooth and rough particles later.

The sign of $N_1$ changes from negative to positive at large $\sigma$ in the DST regime, indicating the presence of dilatancy. Our measurements show that SM colloids display negative $N_1$ values consistently, whereas increasing the RMS roughness shifts the sign change to lower $\phi$ for MR ($\phi = 0.535, 0.55$) and VR colloids ($\phi = 0.50$) (**Fig. 2b**). Many experimental and simulation papers over the years have attributed negative $N_1$ values and a mild increase in $\eta$ to the formation of hydroclusters.[23] More recently, positive $N_1$ values have been attributed to frictional flows and a breakdown of lubrication films.[28, 29] Here, our new simulations support our inference that roughness reduces the criterion for the shear thickening-dilatancy transition due to an increase in the interparticle friction coefficient, $\mu$. **Fig. 3** shows results of dissipative particle dynamics (DPD) simulations[10] on a system of SM particles ($\phi = 0.535$) with tangential interactions that range from completely frictionless ($\mu = 0$) to frictional ($\mu = 1$). Two key points can be taken from **Fig. 3**. First, the value of $\sigma_c$ shifts to lower values as $\mu$ in simulations and roughness in experiments increase (**Fig. 3 Inset**). Because the critical onset stress is directly related to the pairwise force balance, this suggests that tangential interactions reduces the force required to push particles into contact. More significantly, there is a corresponding change in $N_1$ from negative to positive values as $\mu$ increases beyond a critical value, showing that presence of tangential interactions reduces the critical stress for the onset of dilatancy. The simulation trends are in good agreement with our experiments.

A natural follow-up question is: how does roughness contribute to the mechanism of the flow transitions? To address this question, we analyze the Newtonian viscosity, the onset stresses, and the shear thickening power as functions of roughness. **Fig. 4a** shows that $\eta_{r,N}$ diverges more rapidly with increasing $\phi$ for the rougher particles. Our data fit well to the Eilers equation $\eta_{r,N} = \left[1 + 1.5(1-\phi/\phi_{max})^{-1}\right]^2$, where $\phi_{max}$ is the volume fraction at which the Newtonian viscosity diverges. The Eilers equation accounts for packing effects and multi-body hydrodynamics in concentrated suspensions.[34, 35] The $\eta_{r,N}(\phi)$ data for SM and SL colloids fall within the spread measured in previous works,[26, 30, 36, 37] while MR and VR colloids differ significantly. This is not due to uncertainties in $\phi$ for the rough colloids, as we show in **Fig. S2**. Their lower values of $\phi_{max}$ imply that a rough particle tends to occupy an excluded volume larger than that of an equivalent ideal smooth sphere during flow. **Fig. 4b** plots the value of $\phi_{max}$ as a function of particle roughness, along with values for smooth colloids from the literature. Our results show that when the particle roughness is greater than a specific length scale $(B(\psi=0)/a_{eff}^2)^{1/2} \geq 0.069 a_{eff}$), packing becomes increasingly difficult and hence the suspension viscosity diverges at $\phi$ below that of the maximum random close packing of ideal spheres ($\phi =$ 0.64). Interestingly, the value of $\phi_{max}$ for VR colloids is in excellent agreement with the maximum packing reported for frictional particles.[5, 15]

**Fig. 4b** supports our hypothesis that the RMS roughness of particles needs to be greater than the lubrication length scale in order for frictional contacts to contribute to the rheology. This is because the dissipative hydrodynamic forces from squeezing flow are greatly diminished at a particle separation of $h = 0.069\ a_{eff}$ (**Fig. S6**). An analogous transition from the hydrodynamic to the boundary lubrication regime is well known in tribology.[38] Studies of granular packings support our observation that $\phi_{max}$ decreases with increasing roughness, since frictional grains have a lower criterion for isostaticity than frictionless spheres.[39]

**Fig. 4c** shows that as roughness increases, there is a corresponding increase in the shear thickening power as characterized by $\beta$. Solid symbols indicate the appearance of dilatant flows at high $\phi$ for MR and VR colloids. Based on our observations in Fig. 4b, we hypothesize that additional contributions from surface roughness can overcome the combination of hydrodynamic and electrostatic interactions that keep particles apart in Newtonian flow and in mild CST. Thus, we estimate the force it takes to push two PMMA colloids into close contact using the relation $F^* = \sigma_c a_{eff}^2$. It has been shown earlier that the threshold stress scaling is $\sigma_c \propto a^{-2}$ for sterically stabilized PMMA particles, which comes from balancing the lubrication hydrodynamic force and other interparticle forces between the cross sectional area of a particle pair.[30, 40] **Fig. 4d** shows that $F^* = 4.7\ k_BT$/nm for SM colloids at all $\phi/\phi_{max}$, consistent with the range of forces reported for particles grafted with PHSA brushes ($F^* = 2.4 - 6.0\ k_BT$/nm).[30, 41] These force units represent the energy barrier that a pair of smooth particles need to overcome in order to approach more closely to each other on a per nanometer basis. Increasing surface roughness decreases $F^*$ by nearly an order of magnitude. Dilatancy is present at $\phi/\phi_{max} > 0.9$ for MR and VR colloids with $F^* < 0.6$ $k_BT$/nm. Plotting $F^*$ against $\phi/\phi_{max}$ does not collapse the data for different roughness, which suggests that there may be other microscopic mechanisms in addition to packing effects in shear thickening and dilatancy. While granular-like frictional interactions and force networks may be present,[7] elastohydrodynamic lubrication from particle deformation could also cause these changes.[38] For instance, the addition of the crosslinker in our synthesis procedure can result in an increase in the elastic modulus of PMMA by up to 40% at room temperature.[42] However, existing literature suggests that this change in modulus cannot immediately explain the observation of positive $N_1$ values.[10]

The use of well-characterized model colloids with varying degrees of RMS roughness gives us the ability to independently investigate the effects of friction, volume fraction, and shear stress that applies to a broad class of suspensions. According to our picture in **Fig. 5**, lubrication dominates the Newtonian flow of suspensions at low $\phi$, $\sigma$, and roughness. When the RMS roughness increases, lubrication hydrodynamics gradually give way to other microscopic mechanisms and the critical stresses required for shear thickening and dilatancy is progressively lowered. In the case of SM colloids, the onset stress for shear thickening is relatively constant up to $\phi = 0.55$. At the highest $\phi$ and $\sigma$, shear thickening and dilatancy are present because the interparticle forces are sufficiently strong to deform particles[38] or to press them into solid-solid contact.[28] Roughness decreases the onset stress for shear thickening in a manner that is reminiscent of frictional interactions in granular materials, in which the microstructural criterion for mechanical stability is greatly reduced.[39, 43]

Our shear thickening state diagram is consistent with previous studies of dense colloidal suspensions in which roughness is present.[27] This work represents a significant step forward in understanding the full spectrum of steady state rheological response based on the competition between lubrication and roughness, which is made possible by the systematic synthesis of model colloids with increasing particle roughness. Because friction could be a major contributor to the stresses in dense suspensions, our work provides a guiding framework for predicting the rheology of a diverse class of materials with anisotropic particle shapes. For instance, cornstarch is a popular choice in studying jammed materials.[5, 44] Individual granules are rarely spherical in shape and often possess corrugated or irregularly faceted surfaces.[45] Interlocked particles at high $\phi$ can exhibit impeded rotational motion even at rest, which contributes to their excluded volume and frictional interactions during flow. Manipulation of particle roughness and shape thus represent a powerful tool for which the desired thickening response can be built into technology such as liquid body armor, shock absorbers, and robotic grippers.[11, 12] In addition, engineers who work with concentrated slurries can now use particle roughness or geometry as a design parameter to optimize the processing flow conditions for which shear thickening or dilatancy is absent.

**Materials and Methods**

*Synthesis of rough PMMA colloids*

We synthesized size-monodisperse poly(methyl methacrylate) (PMMA) colloids sterically stabilized with a 10-nm layer of poly(12-hydroxystearic acid) (PHSA).[46] A crosslinker molecule, ethylene glycol dimethacrylate (EGDM), was added during the nucleation step of the free-radical polymerization synthesis. We varied the EGDM concentration from 0 - 1.8 wt% of the monomer methyl methacrylate in order to produce a range of particle geometries. After cleaning with pure hexane, we re-suspended the colloids in a mixture of 66:34 v/v% cyclohexylbromide and decalin that provides optimal refractive index and density matching. Electrostatic charges were screened by the addition of 1 mM of tetrabutylammonium chloride. The Debye length of the solvent, $\kappa^{-1}$ = 78 nm, was obtained from conductivity measurements of the solvent, and the zeta potential, $\zeta$, was estimated to be $\leq$ 10 mV. The solvent viscosity at 22°C was measured as $\eta_0$ = 0.0025 Pa·s using a Canon viscometer.

*Surface roughness characterization*

We used an atomic force microscope (AFM) to measure the surface topography of the colloids. After re-dispersing particles in pure hexane, we spin coated the dilute suspensions (1 wt%) onto an Anotop filter glued on a glass slide to produce an even monolayer for imaging. The filter prevents the particles from rolling during AFM imaging. Dried particles were analyzed using an AFM (MFP-3D, Asylum Research) operating in tapping mode with a silicon cantilever (NCH-20, NanoAndMore) with a spring constant of 42 N/m and an intrinsic frequency of 320 kHz. Measured profiles were fitted to an effective sphere of radius $a_{\text{eff}}$ where the deviation of the profile from the ideal sphere was minimized. Only the top 20% of the AFM profiles were analyzed so as to eliminate regions in which the sides of the AFM tip interfered with measurements. We analyzed the profiles of 1, 4, 12, and 7 particles for SM, SL, MR, and VR geometries respectively. Independent size measurements were performed with scanning electron microscopy (SEM), and we averaged the data from at least 50 particles per batch. These SEM measurements showed that the AFM data generated $a_{\text{eff}}$ values with differences of < 4%.

We determined $\phi$ by the mass ratio of the dry particles to the solvent;[47] the crosslinker was added at sufficiently dilute concentrations such that the density of the PMMA particles was assumed constant. The validity of this assumption was subsequently confirmed by counting the number of fluorescently dyed MR colloids in quiescent image volumes (42.5 × 42.5 × 29.0 μm$^3$) from $\phi$ = 0.30 to 0.55 using confocal laser scanning microscopy (**Fig. S2**).

*Steady-state rheological characterization*

Steady state flow measurements were performed with a stress-controlled rheometer (AR-G2, TA Instruments). Colloidal suspensions were loaded onto a Peltier plate set at $T$ = 25°C. We used a $R$ = 20 mm 2° cone-and-plate geometry to perform the majority of our measurements. The geometry was lowered to the truncation gap while rotating slowly (angular frequency, $\omega$ = 0.1 rad/s) to minimize the formation of bubbles at the interface. A solvent trap filled with cyclohexylbromide and decalin was used to minimize evaporation over the duration of the experiments. The suspensions were presheared at 287 s$^{-1}$ for 1 minute, with the exception of dense suspensions of MR and VR colloids (suspensions with $\phi \geq 0.50$ were presheared at 0.5 s$^{-1}$ for 5 minutes) due to the risk of fracturing the sample. We then applied a unidirectional shear consisting of step-wise $\sigma$ increments from 10$^{-3}$ Pa up to a maximum of 400 Pa to characterize the shear thickening behavior of the suspensions. In addition to $\eta$, we also determined $N_1$ values from axial force measurements, $2F_z = N_1 \pi R^2$. Instrument sensitivity limits for $\eta$ and $N_1$ were determined using poly(ethylene glycol) (5 - 4700 cP) and poly(dimethyl siloxane) (50 cSt) standards, which we marked as grey regions in **Fig. 2**. Stress in the up and down directions allowed a check for flow curve hysteresis in two types of suspensions: SM colloids at $\phi = 0.55$, and MR colloids at $\phi = 0.535$ (**Fig. S4**). We checked for slip in our samples with a $R = 30$ mm 1° steel cone-and-plate geometry (**Fig. S5**).

*DPD simulations of frictionless and frictional flows*

Particle motion in our modified DPD model is governed by sum of 5 pairwise interaction forces acting on each particle ($m_i \frac{d\mathbf{v}_i}{dt} = \sum \mathbf{F}_{ij}^C + \mathbf{F}_{ij}^D + \mathbf{F}_{ij}^R + \mathbf{F}_{ij}^H + \mathbf{F}_{ij}^{Contact}$): terms on the right side of the equation are conservative, dissipative, random, hydrodynamic, and contact interaction forces respectively, where the hydrodynamic and the contact forces act only on colloidal particles, and the conservative force only on the solvents. Conservative force is a linearly-decaying repulsive function ($\mathbf{F}_{ij}^C = a_{ij} \omega_{ij}^C (\mathbf{r}_{ij}) \mathbf{e}_{ij}$) that defines the inter-particle pressure between the solvent particles. The random ( $\mathbf{F}_{ij}^R = \sigma_{ij} \omega_{ij}^R (\mathbf{r}_{ij})(\Theta_{ij} / \sqrt{\Delta t}) \mathbf{e}_{ij}$ ), and the dissipative ($\mathbf{F}_{ij}^D = -\gamma_{ij} \omega_{ij}^D (\mathbf{r}_{ij})(\mathbf{v}_{ij} \cdot \mathbf{e}_{ij}) \mathbf{e}_{ij}$) forces form the canonical ensemble and satisfy the Fluctuation-Dissipation theorem, using a random function, $\Theta_{ij}$, of zero mean value and unit variance. In these equations $\Delta t$ is the time step, $\sigma_{ij}$ is the strength of the thermal fluctuations, $\mathbf{v}_{ij} = \mathbf{v}_i - \mathbf{v}_j$ is the relative velocity of interacting particles, and $\gamma_{ij}$ is the dissipation strength. Thus the parameters together define the temperature in the system as $\sigma_{ij}^2 / 2\gamma_{ij} = k_B T$. Weight functions used in these equations are defined as $\omega_{ij}^C(\mathbf{r}_{ij}) = \omega_{ij}^R(\mathbf{r}_{ij}) = [\omega_{ij}^D(\mathbf{r}_{ij})]^{0.5} = \omega_{ij}(\mathbf{r}_{ij}) = 1 - (r_{ij}/r_c)$ vanishing at the cut off distance, $r_c$. In addition to these forces, the lubrication potential, $\mathbf{F}_{ij}^H = -(3\pi\eta_0 a^2 / 2h_{ij})(\mathbf{v}_{ij} \cdot \mathbf{e}_{ij})\mathbf{e}_{ij}$, is introduced for the colloidal particles. The pair drag in squeeze mode lubrication diverges as, $(3\pi\eta_0 a^2 / 2h_{ij})$, at the contact ($h_{ij} = 0$) and thus has to be regularized for small distances ($10^{-4} a$).

In this scheme the contact force, $\mathbf{F}_{ij}^{Contact}$, is adapted from the frictional model by Seto and coworkers:[29] Normal contact ($\mathbf{F}_{N,ij}^{Contact} = k_n h_{ij} \mathbf{e}_{ij}$) acts along the center-center line of interacting particles when $h_{ij} < 0$, and frictional contact ($\mathbf{F}_{T,ij}^{Contact} = k_t \xi_{ij}; \mathbf{T}_{T,ij}^{Contact} = a\mathbf{e}_{ij} \times \mathbf{F}_{T,ij}^{Contact}$) acts tangentially and following Coulomb's friction law ($|\mathbf{F}_{T,ij}^{Contact}| \leq \mu |\mathbf{F}_{N,ij}^{Contact}|$). In these equations $k_n$ and $k_t$ are the normal and tangential spring constants, $\mu$ is the friction coefficient, and $\mathbf{T}_{T,ij}^{Contact}$ is the torque resulting from the tangential displacement, $\xi_{ij}$. If the Coulomb's law is fulfilled, the contact is static and the tangential displacement is updated based on the tangential (relative) velocity of particles ( $\xi_{ij} = \mathbf{V}_{ij}^T dt$ ); Otherwise ( $|\mathbf{F}_{T,ij}^{Contact}| > \mu |\mathbf{F}_{N,ij}^{Contact}|$ ), the sliding contact dictates that $\xi_{ij} = \frac{\mu}{k_t}|\mathbf{F}_{N,ij}^{Contact}|$.

Pressure and stress tensors are calculated from $\mathbf{\Sigma} = -\mathbf{\sigma} = \frac{1}{V}\left\{\sum_{i=1}^{N} m_i \mathbf{v}_i \otimes \mathbf{v}_i + \sum_{j>i}^{N}\sum_{i=1}^{N-1} \mathbf{r}_{ij} \otimes \mathbf{F}_{ij}\right\}$, where $m_i$ is the mass of particle $i$, and $V$ is the volume of the calculation cell. In practice, simulations of suspensions at dimensionless temperature of $k_B T = 1$ with colloidal particles of $a = 1$ and mass of $m = 4/3\rho\pi a^3$ were performed on 1000 colloidal and 18000 solvent particles, with the time-step size of $\Delta t = 5\times 10^{-6} r_C \sqrt{m/k_B T}$. The conservative, dissipative and random parameter was set at $a_{ij} = 25$, $\gamma_{ij} = 50$, and $\sigma_{ij} = 10$ respectively.

**Acknowledgements**

We thank J. Swan, W. Poon, S. Risbud, and A. Hollingsworth for advice and discussion. This work is supported in part by the National Science Foundation (NSF CBET 1232937), a Rackham Predoctoral Fellowship (University of Michigan), and the US Army Research Office through a MURI grant (Award W911NF10-1-0518).

**Author contributions**

L.C.H. conceived and designed the study, synthesized and characterized the colloidal particles, performed all experiments, analyzed data, and interpreted the results. S.J. designed, ran, and analyzed the DPD simulations. D.B.V. analyzed roughness parameters from AFM data collected by E.G. L.C.H., S.J., R.G.L. and M.J.S. analyzed the results and wrote the manuscript. All authors commented on the manuscript.

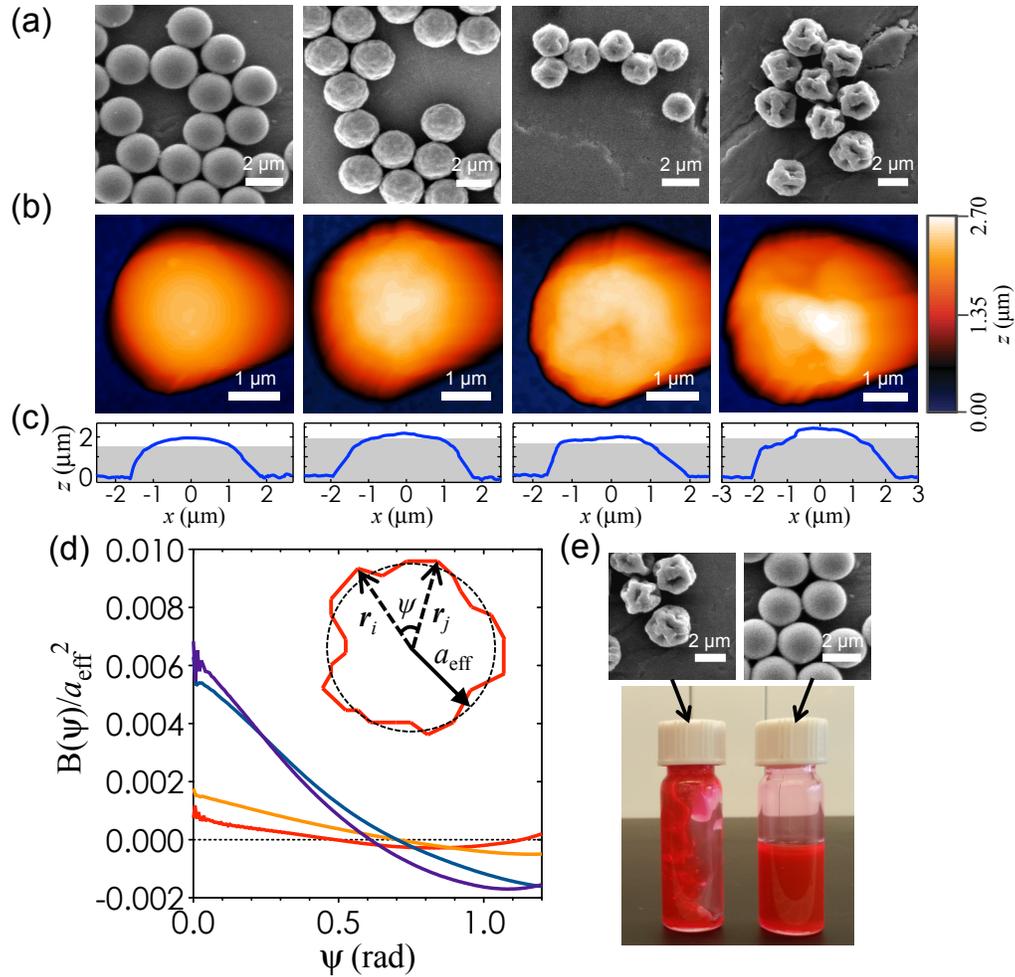

**Figure 1 | Characterization of model colloids with various roughness for suspension rheology.** (a), SEM images of SM, SL, MR, and VR colloids (left to right). (b, c), Representative AFM topography of the colloids, where (b) top-down and (c) side view of the profiles are shown. Grey regions in (c) indicate the bottom 80% of the profile that are limited by cantilever geometry. (d), The top 20% of the AFM topography is used to compute the roughness autocovariance of the different particle geometries (red, SM colloids; orange, SL colloids; blue, MR colloids; purple, VR colloids), where the *y*-intercept indicates the squared RMS roughness. Inset: Schematic of a representative rough particle where the effective radius $a_{eff}$ is found by fitting the radial AFM profile to an ideal sphere. (e), Images of fluorescently dyed MR (left) and SM (right) colloids ($\phi$ = 0.52) in vials that were imaged immediately after turning them upside down. Gravitational stresses exerted on the suspensions of MR colloids caused them to exhibit finger-like structures as they flowed slowly to the bottom of the vial, whereas SM colloids flowed uniformly.

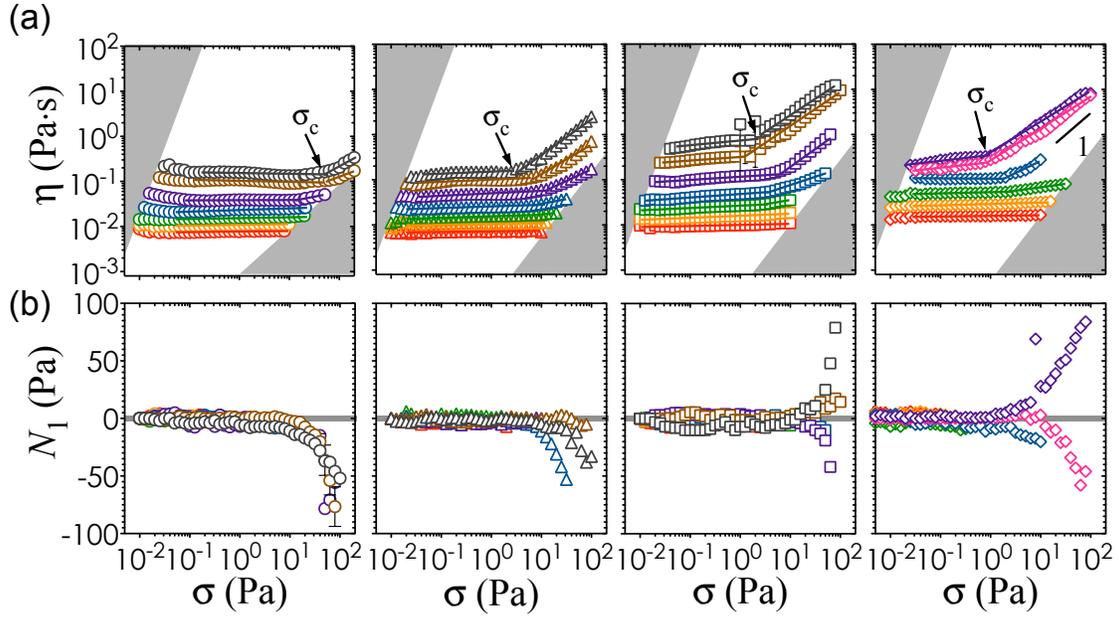

**Figure 2 | Effect of particle roughness on suspension rheology.** (a), Steady state viscosity and (b), $N_1$ flow curves of (left to right) SM, SL, MR, and VR colloidal suspensions (red, $\phi = 0.30$; orange, $\phi = 0.35$; green, $\phi = 0.40$; blue, $\phi = 0.45$; pink, $\phi = 0.48$; purple, $\phi = 0.50$; brown, $\phi = 0.535$; grey, $\phi = 0.55$). Solid lines in (a) are power law fits to the data. Grey regions in (a) indicate instrument sensitivity limits on the left and inertial/fracture effects on the right, while grey regions centered about $N_1 = 0$ Pa in (b) indicate instrument sensitivity limits. Error bars in (a) and (b) are standard deviations from three independent upward stress sweeps.

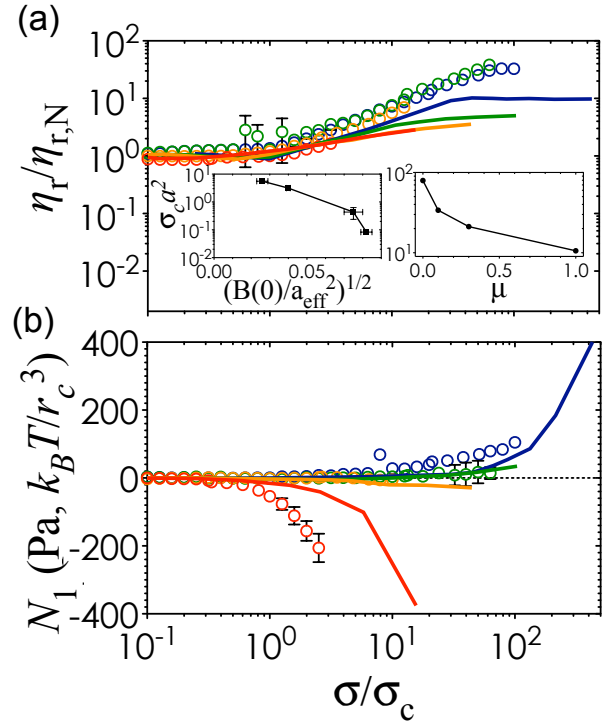

**Figure 3 | Simulated and experimental flow curves of frictionless (smooth) and frictional (rough) particles.** (a), Normalized steady state viscosity and (b), $N_1$ values of colloidal suspensions. Both plots are generated for suspensions at $\phi = 0.535$ (except for VR colloids plotted at $\phi = 0.50$). Data from experiments are plotted as open circles for the four types of roughness tested (red, SM; orange, SL; green, MR; blue, VR) and data from DPD simulations are plotted as solid lines for DPD particles with varying friction coefficients (red, $\mu = 0$; orange, $\mu = 0.1$; green, $\mu = 0.3$; blue, $\mu = 1$). Inset in (a) show the critical onset stress as a function of roughness in experiments (left) and as a function of $\mu$ for DPD particles (right). Error bars represent standard deviations from three independent measurements.

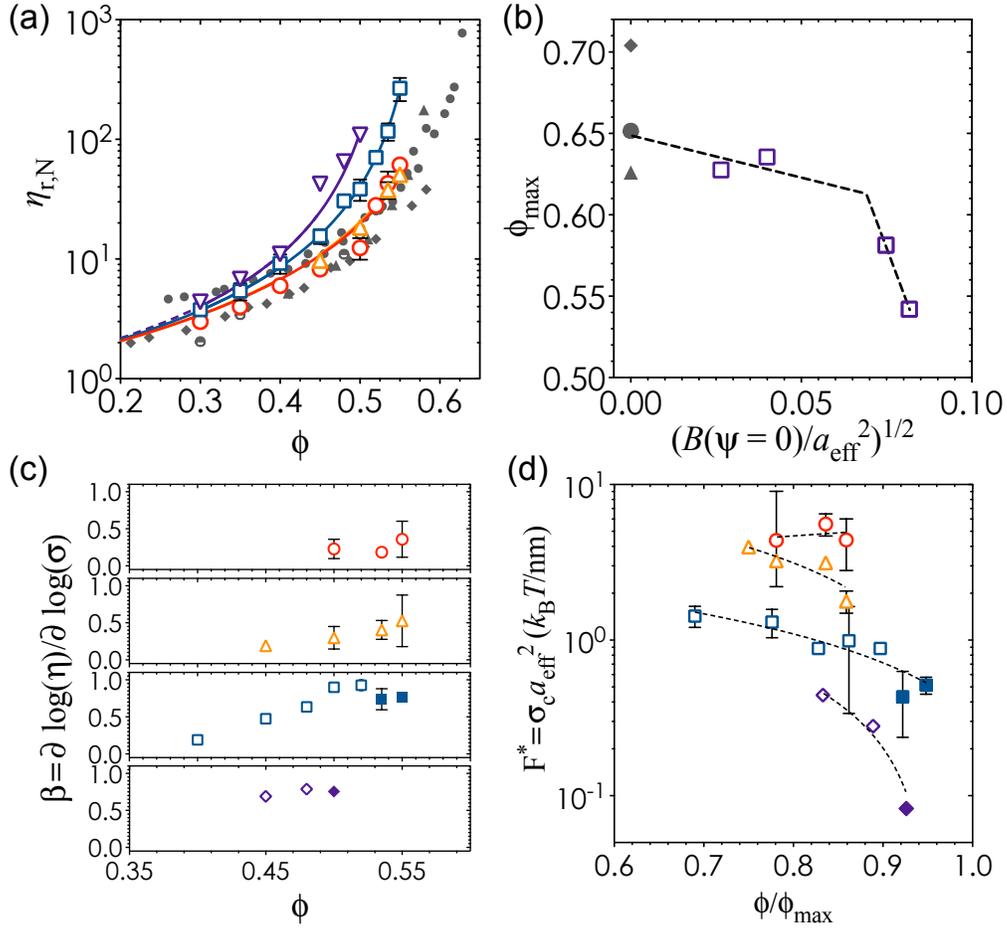

**Figure 4 | Characterization of suspension properties as a function of roughness.** (a), The relative Newtonian viscosity plotted against $\phi$ for SM (red open circles), SL (orange open triangles), MR (blue open squares), and VR (purple open diamonds). Solid lines are fits with Eilers model. Grey filled symbols are data for sterically-stabilized PMMA[30, 36, 37] and for silica colloids.[26] (b), Maximum packing fraction obtained from the Eilers fit plotted against the RMS roughness (open purple squares), where literature data of smooth colloids are included[26, 30, 36, 37] (grey filled symbols). Dashed lines guide the eye. (c), Shear thickening power for suspensions at different $\phi$. Data for different roughness are plotted separately to increase visibility. (d), Force required to push particles into contact plotted against $\phi/\phi_{max}$. Dotted lines guide the eye. In (c) and (d) the symbols follow the same legend as in (a), and filled symbols indicate flow curves in which dilatancy ($N_1(\sigma) > 0$) was observed. Error bars represent standard deviation from three independent measurements.

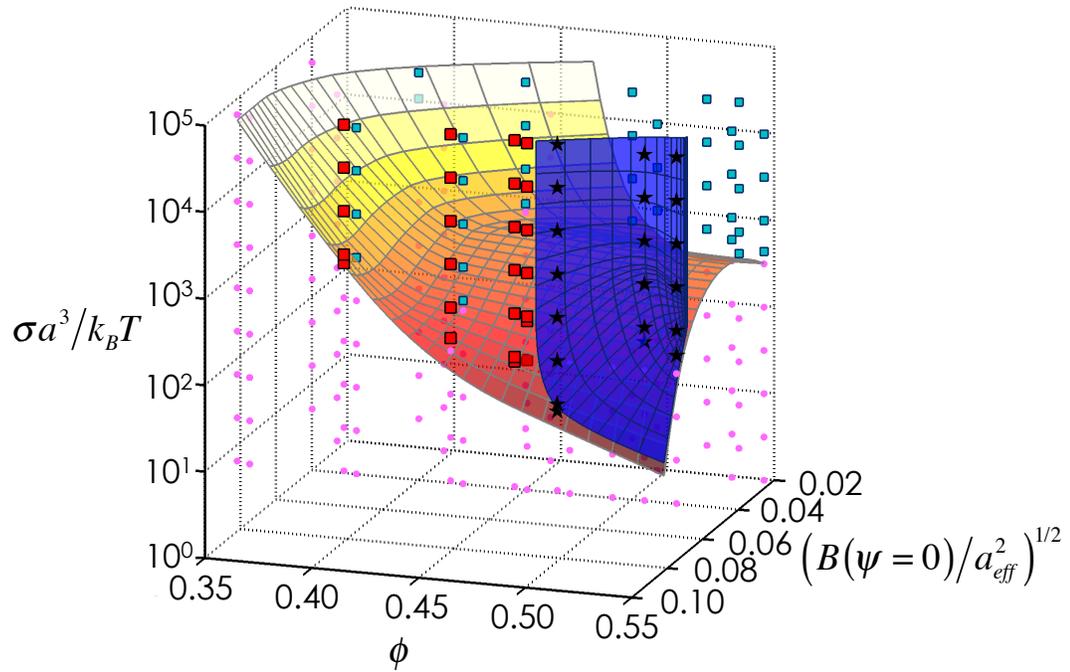

**Figure 5 | A 3D rheological state diagram for rough colloids in sheared suspensions.** This state diagram shows the transition from Newtonian behavior to shear thickening and dilatancy (pink filled circles, Newtonian flows; green filled diamonds, CST ($0.1 \leq \beta \leq 0.65$); purple filled diamonds, DST ($0.65 \leq \beta \leq 1.0$); black filled stars, dilatant flows) as a function of the applied stress, the particle roughness, and the suspension volume fraction. Surfaces indicate the onset of shear thickening (yellow-orange) and dilatancy (dark blue).

# Supplementary Information

## A rheological state diagram for rough colloids in shear flow

Lilian C. Hsiao[1,*], Safa Jamali[2], Daniel J. Beltran-Villegas[3], Emmanouil Glynos[4,§],

Peter F. Green[4], Ronald G. Larson[3] & Michael J. Solomon[3]

**Affiliations:**

[1] Department of Chemical and Biomolecular Engineering, North Carolina State University, Raleigh, NC 27695, USA.

[2] Department of Chemical Engineering, Massachusetts Institute of Technology, Cambridge, MA 02139, USA.

[3] Department of Chemical Engineering, University of Michigan, Ann Arbor, MI 48109, USA.

[4] Department of Material Science and Engineering, University of Michigan, Ann Arbor, MI 48109, USA.

[§] Present address: Institute of Electronic Structure and Laser, Foundation for Research and Technology-Hellas, Crete, Greece.

[*] Corresponding author. e-mail: lilian_hsiao@ncsu.edu.

# Table of Contents

## Synthesis of rough PMMA colloids

Sterically-stabilized PMMA colloids are synthesized in-house via a free radical dispersion polymerization reaction.[1-3] All chemicals are purchased from Sigma-Aldrich unless otherwise specified. First, poly(12-hydroxystearic acid) (PHSA) is grafted onto a PMMA backbone containing glycidyl methacrylate (GMA) anchors.[4] The resultant comb copolymer (PHSA-GMA-MMA) is added with the thermally activated initiator (2-azobisisobutyronitrile, AIBN) to a mixture of dodecane and hexane. The reaction vessel is connected to an Alihn condenser with cooling water and purged with nitrogen gas. Once it is heated to 80°C, the inhibitor-removed monomer methyl methacrylate (MMA) is added. When the reaction begins, a crosslinking agent ethylene glycol dimethacrylate (EGDM) is added using a syringe pump at a rate of 500 μL/min. The crosslinker is thought to induce the microphase separation of co-polymerized oligomers that precipate out of the mixed solvent to form nucleation sites for further growth into primary PMMA particles over 2 hours[5] (**Fig. S1a**). Controlling the ratio of EGDM to MMA allowed us to tune the roughness of the resultant colloids. Some examples of these morphologies are shown in **Fig. S1b** and their reaction conditions are listed in **Table S1**. Particles are cleaned by multiple washes with hexane and stored dry until further use.

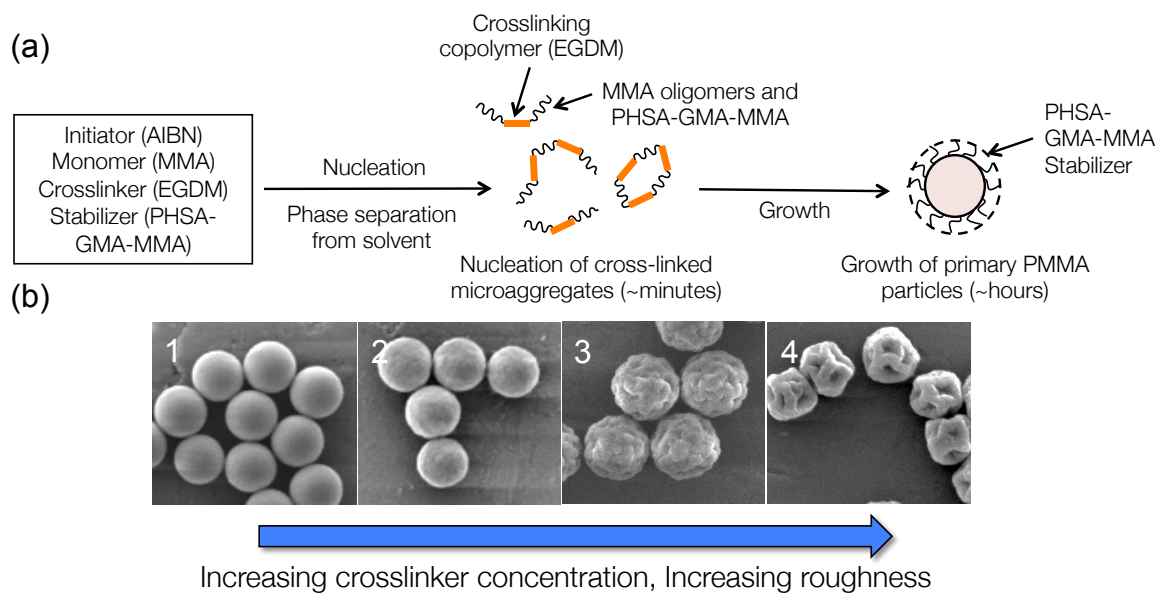

**Figure S1 | Schematic of free-radical polymerization to synthesis rough PMMA colloids.** (a), Reaction pathway showing the basic mechanism of nucleation and growth from the monomer MMA with a crosslinking reagent EGDM added to induce surface roughness. (b), SEM images of sterically-stabilized PMMA colloids with different surface roughness by controlling the ratio of EGDM to MMA during the nucleation step of the polymerization reaction. The reaction conditions for particle batches 1 to 4 are given in Table S1.

**Table S1. Concentration of EGDM and MMA used to make rough PMMA colloids**

| Particle batch # | [EGDM]/[MMA] (v/v%) | $a_{\text{eff}}$ |
|---|---|---|
| 1 | 0.40 | 2.71 μm ± 3% |
| 2 | 0.63 | 2.67 μm ± 2% |
| 3 | 1.33 | 3.53 μm ± 4% |
| 4 | 1.80 | 2.78 μm ± 6% |

## Experimental conditions

In order to conduct reproducible rheological measurements on dense suspensions, large quantities of particles are needed in each run. As such, multiple batches of colloids with the same type of surface morphology were synthesized as shown in **Table S2**.

**Table S2. Range of experimental conditions used in this study**

| Geometry | $2a_{\text{eff}}$ (μm) | $(B(\psi = 0)/a_{\text{eff}}^2)^{1/2}$ | Range of $\phi$ |
|---|---|---|---|
| Smooth | 2.27 μm ± 5%<br>1.60 μm ± 4% | 0.026 ± 0.003 | 0.30 to 0.55 |
| Slightly rough | 2.55 μm ± 2% | 0.040 ± 0.002 | 0.30 to 0.55 |
| Rough | 1.95 μm ± 5%<br>2.06 μm ± 4%<br>2.47 μm ± 5%<br>1.91 μm ± 5% | 0.075 ± 0.005 | 0.30 to 0.55 |
| Very rough | 2.78 μm ± 6% | 0.082 ± 0.003 | 0.30 to 0.50 |

## Calculation of the particle volume fraction with rough colloids

The accurate computation of volume fraction is particularly important in concentrated suspensions where colloids encounter caging constraints due to being in close vicinity of their nearest neighbors.[6] Experimentally, it is simple to estimate a mass-based volume fraction $\phi_W$ by taking the mass ratio of the particles to the solvent because of density matching. To verify that $\phi_W$ represents the true volume fraction of the particles, we use confocal laser scanning microscopy to capture 3D image volumes of fluorescent SM and MR colloids suspended at $\phi_W$ between 0.30 and 0.54 in a mixture of CHB and decalin (66/34 v/v%, **Fig. S2a**). The effective volume fraction $\phi_{CLSM} = \dfrac{4/3 \pi a_{\text{eff}}^3 N_p}{V_{\text{box}}}$ is computed from the total number of particles $N_p$ found in the image volume $V_{\text{box}}$, along with the effective radius of the particles $a_{\text{eff}}$ as obtained from the SEM images. **Fig. S2b** shows that $\phi_W$ is good agreement with $\phi_{eff}$ across the range of volume fractions used in this study.

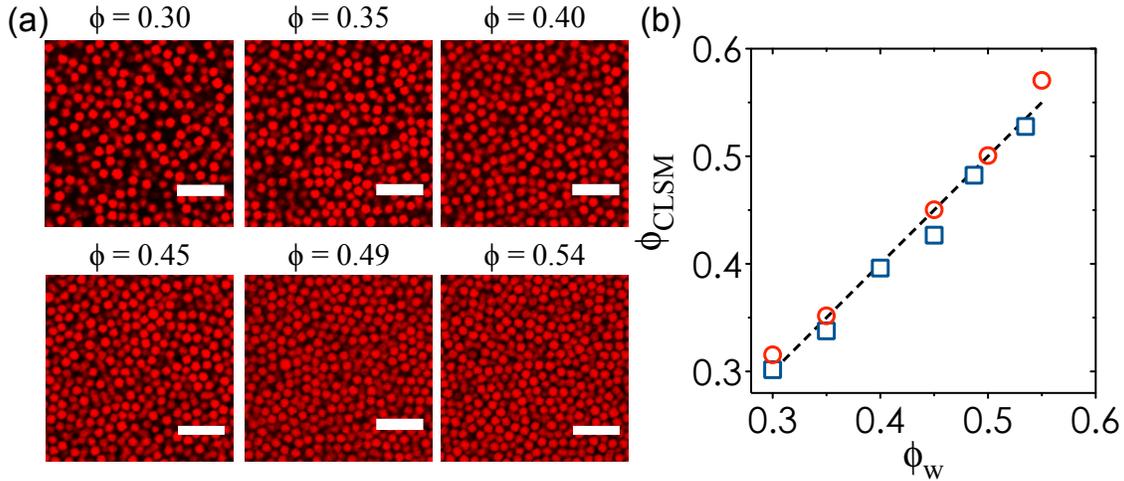

**Figure S2 | Computing particle volume fraction of rough colloids.** (a), Representative confocal laser scanning microscopy images of MR colloids at different $\phi_W$. Scale bars = 10 μm. (b), Particle counting was performed on 3D image stacks of the SM (red open circles) and MR (blue open squares) colloids in order to verify that the mass-based volume fraction ($\phi_W$) matches the particle counting-based volume fraction ($\phi_{CLSM}$). Dashed line indicates $\phi_W = \phi_{CLSM}$.

## Effect of swelling on the particle volume fraction

PMMA colloids have previously been reported to swell in certain organic mixtures which are good solvents for MMA. This swelling could result in errors in particle volume fractions, which would affect our data interpretation because of the divergence in viscosity as $\phi$ approaches random close packing ($\phi_{rcp}$ = 0.64). We investigate the effect of suspending PMMA colloids in the mixture of CHB and decalin used in this study by monitoring changes in particle size over ~ 20 hours. The experiments are conducted with colloidal gels consisting of smooth and MR colloids. Gelation is induced with the addition of small polystyrene molecules (MW = 900,000 g/mol, $R_g$ = 41 nm as determined in previous work from our group[2]) at a concentration of c/c* = 1.0, where c* is the overlap concentration of the polystyrene. 1 μM of TBAC is added to the solvent to provide charge screening. 3D image volumes (53.0 × 53.0 × 25.1 μm$^3$) with voxel dimensions of 0.104 × 0.104 × 0.104 μm$^3$ are captured using confocal laser scanning microscopy. Because particles are in direct contact with each other in a colloidal gel network, we measure the radial distribution function $g(r)$ and take the first peak in $g(r)$ as the effective particle diameter. Changes in the particle radius $a$ are monitored for over ~20 hours. **Fig. S3** shows that un-crosslinked, smooth colloids swell up to 4.0% of their initial size at t = 0, whereas crosslinked MR colloids swell to a lesser extent (2.7%). The uncertainty in the value of $\phi$ is thus 2.3% in this study, with smooth colloids exhibiting greater swelling than any of the rough colloids due to a lack of crosslinking.

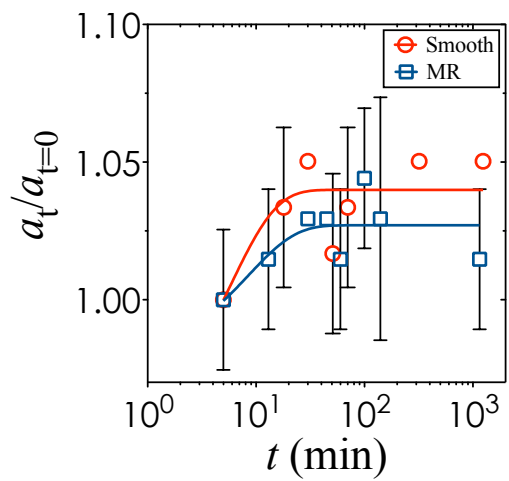

**Figure S3 | Changes in particle radius due to swelling in solvent.** The particle radius, $a_t$, is monitored up to $t = 20$ hours using colloidal gels generated from smooth and MR colloids. The value of $a_t$ is obtained from the first $g(r)$ peak found from processing 3D image volumes of the gels. Error bars represent standard deviations from three independent measurements within the same sample.

## Hysteresis and slip in concentrated suspensions

We observe hysteresis when performing upwards and downwards steady state stress sweeps on MR colloids as shown in **Fig. S4**. However, smooth colloids do not show any hysteresis. This difference suggests that the microscopic mechanisms that drive the shear thickening of smooth and rough colloids are different.

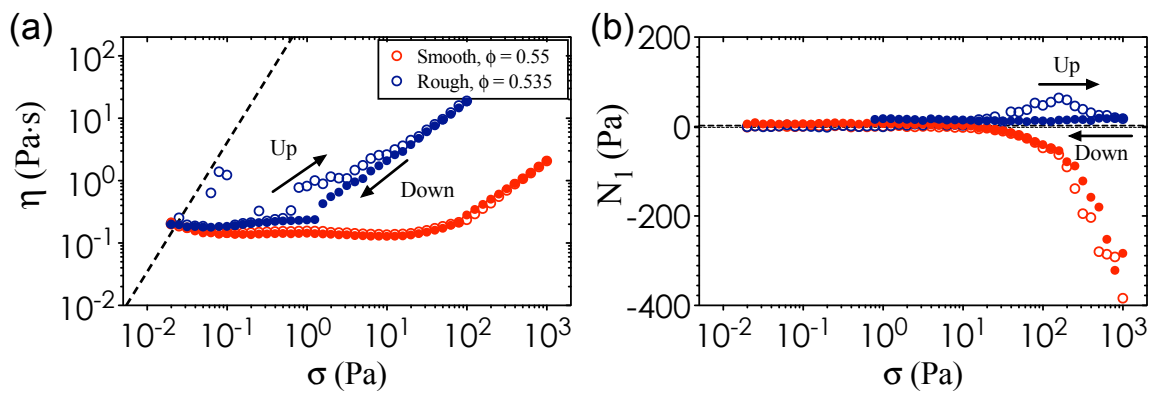

**Figure S4 | Hysteresis of flow curves at high volume fractions.** Upwards (open symbols) and downwards (filled symbols) steady state stress sweeps on suspensions of smooth (red, $\phi = 0.55$) and MR (blue, $\phi = 0.535$) colloids. (a), The steady state viscosity and (b) the normal stress differences are shown for both cases. Dashed lines represents the instrument sensitivity limits.

Slip is also a potential source of uncertainty in concentrated suspensions. We check for the absence of slip by performing steady state upwards stress sweeps on a dense suspension of MR colloids ($\phi = 0.535$) using two different types of cone-and-plate geometries on the rheometer. The results in Fig. S5 show that $\eta$ and $N_1$ values do not differ significantly (to within experimental uncertainties) when the two types of geometries are used. Thus, we conclude that slip does not present a significant source of error in our data interpretation.

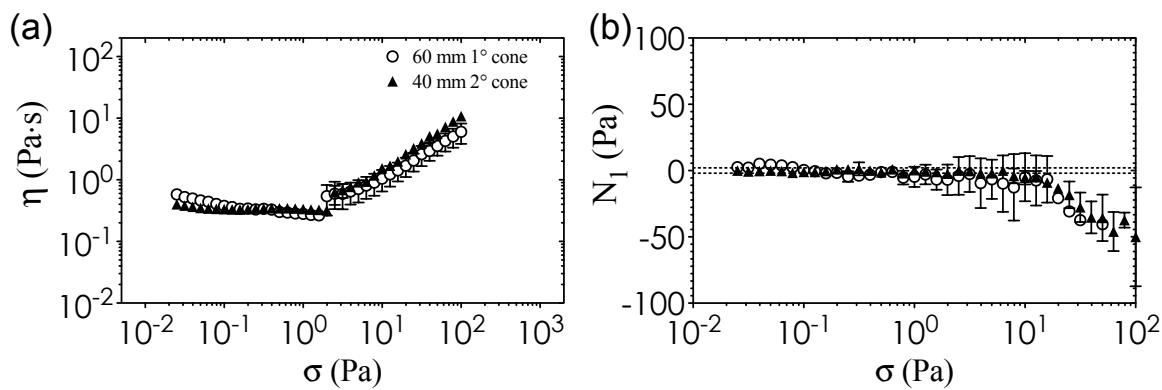

**Figure S5 | Slip in concentrated suspensions.** We checked for slip in a dense suspension of MR colloids ($\phi = 0.535$) using a $R = 30$ mm 1° cone-and-plate (open circles) and a $R = 20$ mm 2° cone-and-plate geometry (filled triangles). The viscosity and N1 flow curves are shown. Error bars represent standard deviation for two independent measurements on the $R = 30$ mm geometry and three independent measurements on the $R = 20$ mm geometry.

## Lubrication hydrodynamic forces from squeezing flow

The dissipative hydrodynamic drag force, $F_{lub}$, diverges at the surface-surface contact of a pair of particles (at particle separation $h = 0$). At high shear rates, the dominant contribution to $F_{lub}$ is from the squeezing of the solvent between the particles ($F_{lub} \sim 1/h$) rather than tangential flows ($F_{lub} \sim \log(1/h)$). In all of the colloidal suspensions tested in our study, the critical shear rate for the onset of shear thickening ranges between 1 s$^{-1}$ to 100 s$^{-1}$. We noted in the main text that when the RMS roughness is at $\sim 0.07$ $a_{eff}$, lubrication effects are diminshed. **Fig. S6** shows that $F_{lub}$ decreases by an order of magnitude between 0.01 $a_{eff}$ and 0.07 $a_{eff}$, suggesting that other microscopic mechanisms that drive shear thickening and dilatancy are greatly enhanced beyond the lubrication length scale.

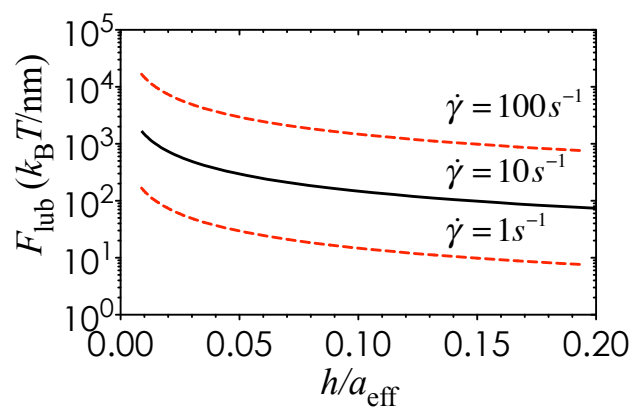

**Figure S6 | Lubrication force from squeezing flow.** The value of $F_{lub}$ is computed using the relation $F_{lub} = 3\pi\dot{\gamma}\eta a^2/2h$ for $\dot{\gamma}$ = 1, 10, and 100 s$^{-1}$ to capture the hydrodynamic forces encountered at the onset condition for all of our shear thickening suspensions.

**Movie of shear thickening suspensions in vials (Movie S1)**

The vials both contain fluorescently dyed PMMA colloids suspended at $\phi = 0.52$ in the solvent described in Materials and Methods. The vial on the left contains smooth colloids, whereas the vial on the right contains MR colloids. Both vials are turned upside down to demonstrate that shear thickening and dilatancy can be tuned using particle roughness.